%
%
\input amstex
\documentstyle{conm-p}
\NoBlackBoxes
\def\IC{{\Bbb C}}
\def\IR{{\Bbb R}}
\def\IZ{{\Bbb Z}}
\def\IH{{\Bbb H}}
\def\II{{\Bbb I}}
\def\CG{{\Cal G}}
\def\CH{{\Cal H}}
\def\CA{{\CA}}
\def\CM{{\Cal M}}
\def\CA{{\Cal A}}
\def\CG{{\Cal G}}
\def\CE{{\Cal E}}
\def\Cm{{\goth m}}
\def\Dsl{\,\raise.15ex\hbox{/}\mkern-13.5mu D}
\def\dsl{\raise.15ex\hbox{/}\kern-.57em\partial}
\def\qft{{quantum field theory}}

\def\h{{\goth h}}
\def\Tr{\hbox{{\rm Tr}}}

\def\Ker{\hbox{{\rm Ker}}}
\def\End{\hbox{{\rm End\,}}}
\def\ad{\hbox{{\rm ad\,}}}
\issueinfo{17}
  {}
  {}
{1998}

\topmatter
\title Unstable Bundles in Quantum Field Theory\endtitle
\author {M. Asorey$^\dagger$, F. Falceto$^\dagger$ 
and G. Luz\'on$^\ddagger$} \endauthor
\leftheadtext{M. Asorey, F. Falceto and G. Luz\'on}%

\address $^\dagger$ Departamento de F\'{\i}sica Te\'orica,
 Facultad de Ciencias,
Universidad de Zaragoza,
50009 Zaragoza, Spain \endaddress

\address $^\ddagger$ Department of Physics, Theoretical Physics,
University of Oxford, 1 Keble Road,
Oxford, OX1 3NP United Kingdom\endaddress

 \thanks \hskip -5pt Based on the Lectures given by M. Asorey at the
 "Conference on Secondary Calculus and Cohomological Physics",
Moscow, Aug. 24-31, 1997 and at the
"Workshop on Algebraic Geometry and Physics", Medina del Campo, 
Sep. 16-20, 1997.
 \endgraf
\hskip -5pt  We acknowledge CICYT for partial financial support under grant
 AEN97-1680.
 \endthanks





\abstract
The relation between connections on 2-dimensional manifolds  and
holomorphic bundles provides a new perspective on the role of
classical gauge fields in quantum field theory in two, three and four
dimensions. 
In particular we show that there is a close relation between unstable 
bundles and  monopoles, sphalerons and instantons.
Some of these classical 
configurations  emerge as nodes of quantum vacuum
states in non-confining phases of \qft\ which suggests a 
relevant role for those configurations in the mechanism of quark
confinement in QCD. 
\endabstract

\endtopmatter

\document

\head 1. Introduction\endhead

The role of principal bundles in the theory of gauge fields
is now well established as the kinematical geometric background where 
classical and quantum fluctuations of gauge fields evolve. However, 
it is much less known that holomorphic bundles also play a
fundamental role in the dynamics of the theory.

There are two cases where the theory of holomorphic bundles
proved to be very useful in the analysis of the 
classical dynamics of  Yang-Mills fields.
The first application arises in the construction  
of  self-dual solutions of four-dimensional Yang-Mills equations 
(instantons), i.e. connections which minimize the Yang-Mills functional
in non-trivial bundles.  
 $U(N)$ instantons in $S^4$ are in one--to--one correspondence with
 holomorphic bundles of rank N  on $CP^3$  which
are trivial along the canonical $U(1)$ fibres \cite{29}.
The second application occurs in 2-D Yang-Mills theory,
where the Yang-Mills functional becomes an almost
equivariantly  perfect Morse function in the space of 2-D
connections \cite{8}. In particular, there are solutions of the
classical Yang-Mills  equations  in every stratum of the
corresponding moduli space of holomorphic bundles.

However, in quantum field theory  all field configurations
are necessary for the formulation of the theory. The quantum states
in the Schr\"odinger representation are functionals on the space of
classical gauge field configurations. The relevance of classical
field configurations for the low energy dynamics of the theory
can be derived from the existence of critical points of the 
modulus of vacuum functionals. In particular the existence of
nodal configurations for the vacuum functional has interesting
consequences on the degeneracy of the quantum vacuum and the
existence of phase transitions.

In this paper we point out that a common feature of all 
classical configurations that appear as nodal points of 
quantum vacua of gauge theories is their relation with 
2-dimensional unstable bundles. Holomorphic bundles also
appear in string theory and F-theory, although we shall not
discuss here these very interesting recent applications
(see \cite{13}\cite{14} for details).

\head 2. Unstable bundles.\endhead

Let $\Sigma$ be a  compact Riemann surface  and let us
recall the definition of holomorphic vector bundles.

\definition{Definition 2.1} A {\it holomorphic  bundle} is a  
vector bundle $E(\Sigma,\IC^N)$ with holomorphic transition functions. 
\enddefinition

The transition functions of a holomorphic bundle
$g_{ij}(x)\in GL(N,\IC)$ satisfy, besides the compatibility
conditions $$
\align 
g_{ii}(x)=\II_N \qquad &{\hbox{\rm  for}}\  x\in U_i\\
g_{ij}(x)\  g_{jk}(x)\   g_{ki}(x) =\II_N  \quad &{\hbox{\rm for}}\  
x\in U_i\cap U_j\cap U_k,
\endalign
$$
 the holomorphic condition
$$ \partial_{\bar z} g_{ij}(x)=0 \qquad {\hbox{\rm  for}}\ x\in 
U_i\cap U_j.$$

An hermitean structure $\h$ can be defined on the same vector
bundle $E$ by unitary transition functions, i.e. $u_{ij}(x)
=h_i^{-1}g_{ij}h_j\in
U(N)$.

\definition{Definition 2.2} Given an 
hermitean structure structure $\h$ in
$E(\Sigma,\IC^N)$ a connection $D:\Gamma(E)\to  \Lambda_1\otimes
\Gamma(E)$ is said  to be {\it unitary} or compatible with $\h$ if
$$ d {\h}(\sigma_1,\sigma_2)= {\h}(D\sigma_1,\sigma_2) + 
{\h}(\sigma_1,D\sigma_2)$$
for any pair of local sections $\sigma_1,\sigma_2$ of $E$.
\enddefinition

\definition{Definition 2.3} For any given hermitean structure
 ${\h}$ of the holomorphic bundle
$E(\Sigma,\IC^N)$ there exists a unique linear connection $D$ on $E$
compatible with $\h$ which  satisfies 
$D_{\bar z} \sigma=0$ for
any local holomorphic section $\sigma$ of $E$. Such a connection
is called the {\it canonical connection} of $E$ with respect to
${\h}$. \enddefinition

In local coordinates $D_{\bar z}=\partial_{\bar z} + { h}^{-1}
\partial_{\bar z} { h}$ 
where ${h}:U\to GL(N,\IC)$ are the coordinates of a given local 
holomorphic frame in a local unitary trivialization of $E$. Notice 
that the local expression does not depend on  the choice of such a 
frame and only depends on the local trivialization of $E$ and the 
holomorphic structure \cite{17}.

Conversely, given a unitary connection $D$ on $E$ with respect to 
the hermitean structure ${\h}$ there exists a unique holomorphic
structure on $E$ whose canonical connection is $D$. This follows from
the fact that the local sections $\sigma$
satisfying the condition $D_{\bar z} \sigma=0$ define a  
holomorphic bundle on $E$, and can be summarized in the following
terms.

\proclaim{Proposition 2.1}
There is a one-to-one correspondence between the space of holomorphic
vector bundles of rank N over a compact Riemann surface $\Sigma$ and
the space of $U(N)$ unitary connections $\CA$ over $\Sigma$ with
respect to a given  hermitean structure $\h$ of the corresponding 
vector bundle.
\endproclaim

In four
dimensions there is a similar correspondence, but in such a case the  
canonical connections associated to   holomorphic bundles must be 
selfdual. In two dimensions there is no  constraint on the associated
unitary connections.

According to Mumford \cite{24} there is a particular class of
holomorphic vector bundles which is distinguished by its stability
properties:  the class of stable vector bundles.

\definition{Definition 2.4} Given a holomorphic bundle over $\Sigma$ 
the ratio $\mu(E)= c(E)/r(E)$ of its first Chern class $ c(E)$ by its 
rank $ r(E)$ is called the slope of $E$.
\enddefinition

\definition{Definition 2.5} A holomorphic vector bundle E
is {\it stable} if there is no holomorphic sub-bundle $F$ with
higher or equal slope, i.e. $\mu(F)<\mu(E)$. If there is no
sub-bundle $F$ with higher slope ($\mu(F)\leq\mu(E)$), $E$ is said 
to be {\it semi-stable}, and
if there exists a sub-bundle $F$ with higher 
slope ($\mu(F)>\mu(E)$), $E$ is said to
be {\it unstable}.
\enddefinition

For genus $g>1$ the space of stable bundles is a dense subset of
the space of holomorphic bundles
and its moduli space $\Cm_{\hbox{\sevenrm s}}$ is a smooth manifold.
These and other  nice features of stable bundles focused the
interest on their study. However,  in quantum
field theory  unstable bundles are acquiring a
prominent role in recent applications.

Let us consider some examples which  appear in the physical
applications

\example{Example 2.1}
There is always a trivial example of holomorphic bundle.
It corresponds to $E=\Sigma\times {\IC}^N$ with constant sections
as holomorphic sections. Any basis of $\IC^N$ defines a global
frame of holomorphic sections that does not require non-trivial
transition functions. 
$E$ is  a semi-stable vector bundle.
 The corresponding canonical connection
associated to any hermitean structure is $A_{\bar z}= 
{ h}^{-1}\partial_z { h}$ with ${ h}:\Sigma
\to GL(N,\IC)$ globally defined on $\Sigma$. This connection
corresponds to a classical vacuum of Yang-Mills theory.
\endexample

\example{Example 2.2}
A non-trivial holomorphic bundle can be associated to the
abelian magnetic monopole.

 Let us consider the complex structure of $S^2$ and the
coordinates induced from $\IR^2$  by the 
stereographic projection. A rank two holomorphic bundle
$E(S^2,\IC^2)$ is given by the transition function
$$g_\pm=
\pmatrix { z} & 0\crcr 0 & 1/z \endpmatrix$$
defined on the overlap $U_+\cap U_-=\{z\ ; \ 1/R<|z|< R\}$ of the two
charts $U_-=\{z\ |\ |z|< R  \}$, $U_+=\{z\ |\ |z|> R  \}$ of $S^2$ 
defined for any radius $R>1$. This vector bundle is   unstable   
because it always admits a line subbundle $L\subset E$ with first 
Chern class $c(L)=1$.

However, with the same holomorphic bundle and hermitean structure
$\h$ we can consider two
different unitary trivializations of E. The first one is given by 
two different unitary trivializations  over $U_-$ and $U_+$ 
connected by the unitary transition function
$$\phi_\pm=\pmatrix { z/|z|} & 0 \crcr
0 & {|z|/ z}\endpmatrix.$$
The corresponding canonical connection, known as
Dirac abelian magnetic monopole,
is  given by
$$A^-_{\bar z}= \phi_-^{-{1}}\partial_{\bar z} \phi_- = {1/2\over
1+|z|^2}\pmatrix z & 0\crcr 0 & -z \endpmatrix \qquad\qquad 
\hbox{\hbox{\rm  for}} \qquad |z|< R$$
with
$$\phi_-=\pmatrix (1+|z|^2)^{{1\over 2}} & 0 \crcr
0 & (1+|z|^2)^{-{1\over 2}}\endpmatrix$$
and
$$A^+_{\bar z}= \phi_+^{-{1}}\partial_{\bar z} \phi_+= {1/2\over
1+|z|^2}\pmatrix {1/\bar z} & 0 \crcr 0 & -{1/\bar z}\endpmatrix
\qquad\qquad \hbox{\hbox{\rm  for}} \qquad |z|> R$$
with
$$\phi_+=\pmatrix {(1+|z|^2)^{1\over 2}/ |z|} & 0 \crcr
0 & {|z|  (1+|z|^2)^{-{1\over 2}}}\endpmatrix$$
in the two charts  $U_+, U_-$ of $S^2$.
Since the unitary trivializations of $E$ were not globally defined
the connection between its two holomorphic trivializations is
defined modulo $\phi_\pm$, i.e.
$$ \phi_+=g_\pm^{-1} \phi_-\phi_\pm .$$ 
The matrices $\phi_+$ and
$\phi_-$ transform the unitary trivializations into 
holomorphic trivializations.

There is, however,  a global
unitary trivialization of $E$ where the
canonical connection reads
$$A_{\bar z}=\varphi_-^{-{1}}\partial_{\bar z} \varphi_-= 
\varphi_+^{-{1}}\partial_{\bar z} \varphi_+= {1\over
(1+|z|^2)^2}\pmatrix {- z} & -1\crcr z^2 & z\endpmatrix,$$
with
$$\varphi_-=\pmatrix z & 1\crcr
{-  (1+|z|^2)^{-1}}&{\bar {z} (1+|z|^2)^{-1}} \endpmatrix ;\,
\varphi_+=\pmatrix 1 & {1/ z} \crcr
{-z (1+|z|^2)^{-1}} &{|z|^2 (1+|z|^2)^{-1}} \endpmatrix$$
Notice that in this case since the unitary trivialization is
globally defined $\varphi_-=g_\pm \varphi_+ .$
It is obvious that in any of the formulations there is line sub-bundle
with first Chern class $c(E)=1$. Thus,    the 
bundle $E$ is unstable.
\endexample

The group $\CG^\IC$ of base preserving linear automorphisms of E
establishes an equivalence relation between different holomorphic
structures. In particular the two holomorphic bundles of  
example 2.2 are equivalent. The unitary automorphism
$$g_-={1\over (1+|z|^2)^{1\over 2}} \pmatrix  - \bar z &1 \crcr
-1 & -z  \endpmatrix \qquad\quad
g_+={1\over (1+|z|^2)^{1\over 2}} \pmatrix  |z|  &-\bar{z} \crcr
{1/ \bar{z}} & |z| \endpmatrix 
\tag 2.1 $$
establishes the equivalence between the two associated canonical 
connections.

The classification of holomorphic bundles over the sphere $S^2$
was carried out by Grothendieck \cite{16}.  
 Holomorphic bundles  over the torus 
  $T^2$ were classified by  Atiyah \cite{7}. The generalization for
arbitrary genus can be found in Ref. \cite{8} and references
therein.

 The space of holomorphic vector bundles  over 
a Riemann surface, $\CH$, can be identified, by Proposition 2.1,
with the space of unitary 
connections $\CA$ once a hermitean structure $\h$ of $E$ is fixed. 
However, in both spaces there are two very different actions of
automorphism groups. In  $\CH$ there is a   natural action  of
  automorphisms $\CG^\IC$ of $E$ as a complex vector bundle, whereas in
$\CA$ only the subgroup of unitary automorphisms $\CG$ with respect to 
$\h$ have a proper action. The two quotient spaces have a very
different structure, they are  called
moduli space $\Cm$ and orbit space $\CM$, respectively.

However, none of those quotient spaces is a smooth manifold because
the  isotopy group of the automorphism groups actions are not the
same for all orbits. In the case of unitary connections there
 two alternatives to provide the orbit space $\CM$ with an smooth
structure: whether restrict $\CA$ to the space of irreducible
(generic) connections $\CA^\ast$, or consider the whole space of all
connections $\CA$ and restrict the automorphism group $\CG$ to the group of
unitary automorphisms $\CG_0$ of $E$ that leave the fiber over a
given point of $\Sigma$ invariant ({\it pointed automorphisms}). 
In both cases the corresponding orbit spaces, $\CM^\ast=
\CA^\ast/\CG$, $\CM_0= \CA/\CG_0$ are smooth manifolds. In
the case of holomorphic bundles we have not such alternatives
because the variety of isotopy groups is wider. The one-dimensional
subgroup of trivial automorphisms of the form $\hbox{\rm exp}
(i\alpha) \II\in\CG^\IC$ leaves any holomorphic bundle invariant. All isotopy
groups contain this trivial subgroup. But the isotopy groups of
stable, semi-stable and unstable vector bundles are quite different
as the following results point out.

\proclaim{Proposition 2.2}
The only automorphisms which leave a stable bundle  invariant are
the trivial automorphisms of the form $\hbox{\rm exp} (i\alpha) \II$.
\endproclaim  
See Ref. \cite{26}  for a proof.

\proclaim{Proposition 2.3}
The only pointed automorphism which leave a semi-stable bundle 
invariant is the trivial automorphism  $\II$. 
\endproclaim  
\demo{Proof}
Let  $\Phi$ be a pointed automorphism leaving the holomorphic
structure of $E$ invariant. $\Phi$ is, thus, a holomorphic section
of  vector bundle $\End E$ of  endomorphisms of $E$, i.e.
 $\phi\in H^0(M, \End E)$.
We can associate to $\Phi$ an holomorphic section  of
the adjoint bundle $\ad E$, $\xi=\Phi-
{\II}$. By
construction $\xi$ vanishes ($\xi(x)=0$) at the point  $x\in\Sigma$
where $\Phi(x)=\II$. Then, the  holomorphic line sub-bundle
$L_\xi$ of  $\ad E$ induced by $\xi$ must have a positive Chern
class $c(L_\xi)>0$. Now, since $\ad E$ inherits the  semi-stable
character of $E$ (see. e.g. \cite{8} for a
proof) and has $0$ slope, this is not possible unless 
 $\xi\equiv 0$, i.e.  $\Phi= \II$.
\qed
\enddemo

The results point out the differences between 
isotopy groups of the different
types of holomorphic bundles over a Riemann surface.
The isotopy group of stable bundles only contains  trivial
automorphisms, whereas that of semi-stable bundles might contain
some non-trivial automorphisms but they can never leave invariant 
any fiber of the bundle.
The following result shows that the isotopy group of unstable bundles
can  contain pointed automorphisms.

\proclaim{Proposition 2.4}
Any unstable bundle $E$ defined over the sphere 
with $c(E)=0$ is invariant under some
pointed automorphisms. 
\endproclaim
\demo{Proof}
Since $E$ is unstable the adjoint bundle $\ad E$ is also unstable.
There exists, thus, a sub-bundle $F$  of  rank $M$  of $\ad E$  with
$c(F)>0$. The Riemann-Roch theorem establishes that
$$H^0(\Sigma,  F) - H^1(\Sigma,  F)=c(F)+  M.$$
Since $c(F)>0$, the bundle $F$ has more holomorphic sections
than its rank. This implies that there is a linear combination of 
them, $\phi\in H^0(\Sigma, F)$, which vanish at a given point. 
The automorphisms of $E$ generated
by this holomorphic section $\phi$ leave the bundles $F$ and $E$ invariant.
\qed
\enddemo

It follows from the previous discussion that the moduli space 
$\Cm_{\hbox{\sevenrm s}}$ of equivalent 
stable bundles over a Riemann surface  is a
finite dimensional smooth manifold, whereas the orbit space 
$\CM^\ast=\CA^\ast/ \CG$ of all
irreducible connections is an infinite dimensional manifold.
The characterization of stable bundles by the irreducible
(projective) unitary representation of the first homotopy group
of $\Sigma$ due to Narasimhan and Seshadri \cite{26}, 
provides a very useful approach to analyze the geometrical 
structure of the corresponding moduli space $\Cm_{\hbox{\sevenrm
s}}$ \cite{28}\cite{25}.

If we include semi-stable bundles the smooth nature of the
quotient space disappears as they might have larger isotopy groups.
There is, however, a way of associating a smooth manifold to the
corresponding moduli $\Cm_{\hbox{\sevenrm
ss}}$. The quotient space of all semi-stable bundles
by the group of pointed automorphisms has a smooth structure
\cite{11}. The procedure can be extended to include classes of
unstable bundles but it is not possible to find a manifold for the
whole set of holomorphic bundles different from the space $\CH$
itself because the dimension of the isotopy group of unstable
bundles is unbounded above.

Let us illustrate these properties with the examples considered
above.

In the case of Example 2.1, all the constant gauge transformations
leave the trivial  holomorphic bundle invariant, which is 
semi-stable.  The unstable bundle of  example 2.2 has four
holomorphic sections  $\chi_{(\mu)};\, \mu=0,1,2, 3$ in the
adjoint bundle. Indeed the sections $$
\align
\chi_{(3)}^-=&\chi_{(3)}^+=\pmatrix
-1 &0 \crcr 0&1 \endpmatrix \\
\chi_{(k)}^-=&{z^k\over 1+|z|^2} \pmatrix
0 &1 \crcr 0&0 \endpmatrix\qquad\quad  k=0,1,2\\
\chi_{(k)}^+=&{z^k\over 1+|z|^2} \pmatrix
0 &{\bar z}/z \crcr 0&0 \endpmatrix\qquad k=0,1,2
\tag 2.2
\endalign
$$
satisfy that $D_{\bar z} \chi_{(\mu)}=0$ in the first trivialization. 
The section $\chi_{(2)}$ vanishes at $z=0$, 
$\chi_{(0)}$ at $ z=\infty$ and 
$\chi_{(1)}$ at both points.

In the second trivialization these four holomorphic sections 
$\chi_{(\mu)};\, \mu=0,1,2, 3$ of the adjoint bundle read
$$
\align
\chi_{(3)}= & {1\over 1+|z|^2}
\pmatrix
|z|^2-1 &  2 {\bar z} \crcr 2 z & 1-|z|^2
\endpmatrix\\
\chi_{(k)}= & {z^k\over (1+|z|^2)^2} 
\pmatrix
\bar z & -{\bar z}^2 \crcr 1 & - \bar z 
\endpmatrix\qquad\quad k=0,1,2,\\
\tag 2.3
\endalign
$$
and again $\chi_{(2)}$ vanishes at $z=0$, 
$\chi_{(0)}$ at $ z=\infty$ and 
$\chi_{(1)}$ at both points.
Since the holomorphic and hermitean structures are the same, 
and the 
only difference
comes from the choice of different unitary trivializations,
the  sections (2.3) are just the transformed of (2.2) under the
(chiral) complex gauge transformation (2.1).

However, the complete characterization of unstable bundles is quite
complex \cite{8}. The space $\CH$ has a stratified structure. Every
holomorphic bundle has a canonical filtration by subbundles
$0=E_0\subset E_1 \subset E_2\subset\cdots \subset E_r=E$,  
such that the quotients $D_i=E_i/E_{i-1}$ are semi-stable with
 strictly decreasing slopes, $\mu(D_1)>\mu(D_2)>\cdots>\mu(D_r)$.
The different strata of $\CH$ are defined by all the holomorphic
bundles which have the same pairs of integers $(r(D_i)_i,c(D_i));\ 
i=1,\cdots, r$ in their canonical filtrations. In all the strata
there are totally decomposable bundles $E=
D_1\oplus D_2\oplus\cdots\oplus D_r$ which are invariant under
non-trivial automorphisms of the form $U=\hbox{\rm exp} (i\alpha_1)
\II_1\oplus\hbox{\rm exp} (i\alpha_2)
\II_2\oplus \cdots\oplus\hbox{\rm exp} (i\alpha_r)
\II_r$. But, the isotopy group of the different orbits of a given
stratum is not necessarily the same. The codimension of the stratum
associated to a canonical filtration $(r(D_i)_i,c(D_i));\ 
i=1,\cdots, r$ 
$$\sum_{i<j}\big[ c(D_i)r(D_j)-c(D_j) r(D_i) - r(D_i)r(D_j)(1-g)\big]
$$
reflects the fact that the increase of the dimension of the isotopy
group for higher strata is compensated by the increase of the codimension
of the strata modulo the change of dimension of the strata.

\head 3. Quantum Fermions\endhead

The first application of unstable bundles appears in the study of 1+1 
dimensional QCD. The effective action of massless Dirac fermions becomes 
divergent for connections that correspond to unstable bundles,
because  of the existence of zero modes for the Dirac operator
\cite{20}.

\proclaim{Proposition 3.1}
The  determinant of the Dirac operator $\Dsl_E$ vanishes for connections 
associated to
unstable bundles $E$ in any stratum.
\endproclaim
\demo{Proof:}
Since $E$ is unstable there is a subbundle $F$ with $c(F)>0$. By index 
theorem the asymmetry in the number of zero modes of the Dirac operator
$\Dsl$ over the bundle $F$  with positive and
negative chirality is given by
$$\dim \Ker\,\Dsl^+_F - \dim \Ker\,\Dsl_F^- =c(F)>0.$$ In particular this
implies  the existence of non-trivial zero modes of positive
chirality. Since $$0<\dim \Ker\,\Dsl^+_F + \dim \Ker\,\Dsl_F^-
\leq\dim \Ker\,\Dsl^+_E + \dim \Ker\,\Dsl_E^-$$ we conclude the existence of
zero modes of the Dirac operator $\Dsl_E$.
\qed
\enddemo

\head 4. Chern-Simons Theory\endhead

The topological character of the theory implies that the
hamiltonian is identically null $H\equiv 0$. There are, however, two
very stringent constraints on physical states $\Psi (A)$:
i) they only depend   on the $A_{\bar z}$ components of
unitary connections $A$, i.e.
$${\delta\over \delta A_{z}} \Psi (A)= 0,$$
 and ii) they must satisfy the Gauss law 
$$D_{\bar {z}} {\delta\over \delta A_{\bar z}} \Psi \left(A_{\bar
z}\right)= {k\over \pi} \partial_z A_{\bar z} \Psi \left(A_{\bar
z}\right).$$ 
The geometrical meaning of this law becomes more explicit if we 
multiply both sides of the equation by an arbitrary function
$\chi:\Sigma\to {\goth{sl}}(N,\IC)$. Integrating
by parts on the left hand side we obtain
$$-\int_\Sigma \Tr \left( D_{\bar z} \chi \right)
{\delta\over \delta A_{\bar z}} \Psi \left(A_{\bar
z}\right)= {k\over \pi} \int_\Sigma \Tr\, \chi\,\partial_z A_{\bar
z} \Psi \left(A_{\bar z}\right),
\tag{4.1}$$
which encodes the transformation law of the physical states under
infinitesimal chiral  {\it complex}
gauge transformations. The integration of Gauss law, thus,
defines the action of
the group of chiral or {\it complex}
gauge transformations $\CG^{\IC}$ on the physical states 
$\Psi \left(A_{\bar
z}\right)$.
This 
action of $h\in \CG^{\IC}$ on  $A_{\bar{z}}$
is given by
$$
{^h A_{\bar{z}}=h A_{\bar{z}}h^{-1}+
h{\partial_{\bar z}}h^{-1},}
\tag{4.2}
$$
and the isomorphism between ${\CA}$ and  $\CH$
induces an action of $\CG^{\IC}$ on ${\CA}$ which
extends that of ordinary unitary gauge transformations in ${\CG}$.

\proclaim{Theorem 4.1}
 The quantum states of Chern-Simons theory  vanish for all
connections associated to unstable bundles defined on the sphere or the
torus.
\endproclaim
\demo{Proof}
The case of the sphere trivially follows  from the fact that
every holomorphic bundle on $S^2$ is decomposable as a sum of line bundles.
For the torus a more elaborated argument is required.
Let us first consider unstable bundles which  split  into
 two stable bundles,
	$$E=D_+ \oplus D_-$$
with   Chern classes $c(D_\pm)=\pm 1$.  Due to
the low values of the genus of $\Sigma$ and the Chern classes
of $D_\pm$ both stable bundles do admit a canonical 
decomposition of the form
$$D_+=\pmatrix L^{\mu_{1}}_0&X_1^2&X_1^3&\cdots& X_1^r\crcr
		0&L^{\mu_{2}}_0&X_2^3&\cdots& X_2^r\crcr
		0&0&L^{\mu_{3}}_0&\cdots &X_2^r \crcr
		 & \cdots & \cdots & \cdots & \crcr
		0&0&\cdots &\cdots &L^{\mu_{r}}_1 \crcr \endpmatrix \qquad
		D_-=\pmatrix L^{\nu_{1}}_{-1}&Y_1^2&Y_1^3&\cdots &Y_1^s\crcr
			0&L^{\nu_{2}}_0&Y_2^3&\cdots& Y_2^s\crcr
		0&0&L^{\nu_{3}}_0&\cdots &Y_2^s \crcr
		 \cdots & \cdots & \cdots & \cdots & \cdots \crcr
		0&0&\cdots &\cdots &L^{\nu_{s}}_0 \crcr \endpmatrix
$$
where $L^\mu_n$ denotes a line bundle with Chern class $c(E_n)=n$
and holonomy $\mu$. It is trivial
to see that generically (for holonomies in generic relative positions) 
the bundle $D_+\oplus D_-$ do not admit extensions, i.e.
$H^1(D_+\otimes D_-^\ast)=0$. Thus,
the orbits of split bundles $E= D_+\oplus D_-$ define a dense subset 
of the space
of unstable bundles of the corresponding strata.
Moreover, those orbits  define a dense subset in the space of all 
unstable bundles. Now, if we saturate both sides of  Gauss law
(4.1) with the
 holomorphic section $\chi$
of $H^0(\ad E)$ defined by
$$\chi=\pmatrix   r \II_1 &0\crcr
0& -s\II_2,        \endpmatrix,$$
 the left hand side vanishes because $\chi$ is holomorphic
$D_{\bar{z}}\chi=0$  whereas the right hand side becomes proportional to
$$\eqalign{
\int _\Sigma\, {\Tr}\,  \chi\,
\partial_{{z}} A_{\bar{z}} &=\int_\Sigma\, {\Tr} \, \left(\chi\,
 \partial_{{z}} A_{\bar{z}} -  A_z D_{\bar{z}}\chi \right)
=\int_\Sigma\, {\Tr}\chi\,\, F(A)\cr
&=2\pi \left( 
s\, c(D_+)- r\, c(D_-)\right)=2\pi(r+s),}
$$
which is a non vanishing factor. Thus, consistency of Gauss law
requires  
 the vanishing of any physical state   $\Psi$  for those
 connections. By continuity we conclude that the nodal set of  $\Psi$
 include all connections in the adherence of the corresponding strata,
 which proves the theorem, i.e.
 $$\Psi(A_{{\hbox{\sevenrm uns}}})=0.$$
 \qed
\enddemo
This vanishing result is in agreement with the limit of
the explicit expressions of  Chern-Simons states 
for connections associated to semi-stable bundles \cite{15}
\cite{11}.

\head 5. Topologically Massive Gauge Theory \endhead

In this case the theory has propagating degrees of freedom and it is
non-topological. The hamiltonian is given by
$$
{
\IH= -{{g^2}\over 2 }\, \left|\left|{\delta\over\delta {A}}+{{i k\over
4\pi}} \ast{A}\right|\right|^2+{1\over 2{g^2}}||F({A})||^2}
\tag{5.1}
$$
and the physical states $\Psi({A})$ are constrained by  Gauss law
$$
-i D^\ast_{A}{\delta\over\delta {A}}\Psi({A})=
{{k\over 4\pi}}\ast d{A}\  \psi({A}).
\tag{5.2}
$$
\proclaim{Proposition 5.1}
 The quantum states of topologically massive gauge theories vanish
for all connections which are  reducible to unitary connections
with non-trivial Chern class.
\endproclaim
\demo{Proof}
Let  $A$ be a connection which is reducible to a unitary
connection $A_0$ with non-trivial Chern class $c(A_0)\neq 0$.
Because of the hermitean structure of $E$ the connection
 $A$ can be split into
two components $A=A_0\oplus A_1$ with non-trivial Chern
classes, $c(A_0)=-c(A_1)\neq 0$. 
Reducible connections always have a non-trivial isotopy group under
the action of the group of gauge transformations $\CG$. In
particular, the section of  the adjoint bundle
$${
\chi=\pmatrix{s\, c(A_0)}\II_0& 0\crcr
0& r\, {c(A_1)}\II_1
\endpmatrix,}
$$
 generates a one-parametric subgroup 
of automorphisms of $E$ which leave  $A$ invariant.
This follows from the trivial identity $D_{A_{{}}}\chi=0$.
Consequently, 
$${
{\int d^2x\,{\Tr}\, \chi\,
D^\ast_A{\delta\over \delta A } \Psi(
A_{})= 0, }}$$
for any functional $\Psi(A)$, which implies that    
  $$\int_\Sigma\, {\Tr} \, \chi\, d A_{} \, \Psi( A_{ })=0$$ 
for physical states, because of Gauss' law.
This is impossible, since 
$$
\int _\Sigma\, {\Tr}\,  \chi\,
d A =\int_\Sigma\, {\Tr} \, \chi\,
 (d A+D_{A}A) 
=2\int_\Sigma\, {\Tr}\,  \chi \, F(A)=4\pi\left[
s\, c(A_0)^2+ r\, c(A_1)^2 \right]\neq 0,$$
unless the physical state $\Psi$ vanishes for that connection, i.e.
$\Psi(A)=0$. 
\qed
\enddemo

Now, reducible connections do not
exhaust all nodal configurations of physical states. 
The following topological arguments show that any physical
state must have additional nodes on irreducible connections. 

\proclaim{Proposition 5.2}
{ Non-trivial solutions of Gauss law
exist if and only if $k$ is an integer. In such a case these solutions
are in one-to-one correspondence with sections of a line bundle
$\CE_k(\CM^\ast,\IC)$ over the orbit space of irreducible
connections, with non-trivial first Chern class $c_1(\CE_k)\in
H^2(\CM^\ast,\IZ)\equiv \IZ$} \endproclaim
\demo{Proof}
Gauss law  has in this case a simple geometric interpretation in 
terms of the
hermitean $U(1)$ connection $\widetilde{\alpha}_k$ defined over 
the space of irreducible connections $\CA^\ast$
by  the one-form
$$
\widetilde{\alpha}_k(\tau)={{k\over 4\pi}}(\ast
{A},\tau)+  {{k\over 2\pi}} (D^{\ }_{A} G^{\ }_{A}\ast
F({A}),\tau)\qquad\qquad  \forall\tau\in T_A \CA  $$
with $G^{\ }_{A}= (D^\ast_{A}D ^{\ }_{A})^{-1}$.
Actually, the quantum Gauss law condition (5.2) can be
written  as
$$
 D^\ast_{A}\nabla_{\widetilde{\alpha}_k} \psi({A})=
0 
\tag{5.3}$$
with
$$ {\nabla_{\widetilde{\alpha}_k} }= {\delta\over\delta {A}
}+i\widetilde{\alpha}_k,$$ 
which means that the quantum states
are covariantly constant along the gauge fibers with respect to the
$U(1)$ connection $\widetilde{\alpha}_ k$. The existence of
non-trivial solutions of the quantum Gauss condition (5.3) is
possible if and only if the connection $\widetilde{\alpha}_k$ is
trivial  along
the orbits of the group of gauge transformations $\CG$
(modulo its center) \cite{5}\cite{1}.  The curvature
2-form of $\widetilde{\alpha}_k$
$${
\widetilde{\Omega}_k(\widetilde{\tau}, \widetilde{\eta})=-
{{k\over 4\pi}}  (\widetilde{\tau}, \ast\widetilde{\eta})
+{{k\over 2\pi}} (G^{\ }_{A}\ast[\widetilde{\tau},
\widetilde{\eta}], \ast F({A})) }
\tag{5.4}$$
 vanishes for  vectors
$\widetilde{\tau}, \widetilde{\eta}\in T_{A}\CA^\ast$ tangent to
the gauge fibers $\widetilde{\tau}=D^{\ }_{A} \phi$. However,
$\widetilde{\alpha}_k$ is trivial only if the holonomy group
associated to any closed curve contained in a gauge orbit is
trivial. This is only possible if the  projection $\Omega_k$
of the curvature 2-form $\widetilde{\Omega}_k$ to the  space of
irreducible gauge orbits $\CM^\ast=\CA^\ast/\CG$,
$$
{\Omega_k}({\tau}, {\eta})=
\widetilde{\Omega}_k(\widetilde{\tau}^h, \widetilde{\eta}^h)
\tag{5.5}$$
belongs (modulo a factor $2\pi$) to an integer
cohomology class of $\CM^\ast$, i.e.
$$
{\textstyle{1\over 2\pi}} [\Omega_k]\in H^2(\CM^\ast, \IZ).
\tag{5.6}
$$
In equation (5.5) $\widetilde{\tau}^h$ denotes the  horizontal
component  of any tangent vector $\widetilde{\tau} \in T_{A}\CA^\ast$
with projection   ${\tau} \in T_{[A]}\CM^\ast$ which  is
orthogonal to the gauge fiber at ${A}$  with respect to the
canonical $L^2$ product defined in ${\CA}^\ast$ by the Riemaniann metric
of $\Sigma$, 
i.e. $\widetilde{\tau}^h=P_{A}\widetilde{\tau}$, $P_{A}=(I-D^{\
}_{A}G^{\ }_{A}D^{\ast }_{A})$ being the corresponding orthogonal
projector.

The condition (5.6) is satisfied  if and only if the
Chern-Simons charge $k$ is an integer. Because of the triviality of 
$\widetilde{\alpha}_k$ when $k$ is an integer, the action of the
group of gauge transformations $\CG$ can be globally lifted to an
action on the line bundle $\CA^\ast\times \IC $. Gauss 
law implies the
invariance of  quantum states under this action. 
Thus, the
quantum states can be completely characterized by sections of the
line bundle $\CE_k(\CM^\ast,\IC)$  defined by the gauge orbits
$\CE_k=\CA^\ast\times\IC/\CG$ of such an action. It is obvious that
the first Chern class of this bundle $c_1(\CE_k)=
\textstyle {1\over 2\pi} [\Omega_k]$ is non-trivial provided $k\neq
0$
\qed
\enddemo

In the same way, the
connection $\widetilde{\alpha}_k$ of $\CA^\ast\times \IC$ projects down
to a connection ${\alpha_k}$ in $\CE_k$ and the quantum hamiltonian
can  be expressed as an operator
$$
\IH= -{{g^2}\over 2 }||\nabla_{\alpha_k}||^2+{1\over 2
{g^2}}||F({A})||^2 + {k^2 {g^2}\over
8\pi^2} \left(\ast
F({A}), G_{A} \ast F({A})\right)
$$
acting on the sections of $\CE_k$. 


\proclaim{Theorem 5.3} {Every physical state has nodal
configurations on the orbit space of irreducible connections 
$\CM^\ast$}
\endproclaim
\demo{Proof} Since the Chern class of the line
bundle $\CE_k(\CM^\ast,\IC)$  does not vanish
$c_1(\CE_k)=[\Omega_k]/2\pi\neq 0$,  the bundle
$\CE_k(\CM^\ast,\IC)$ is non-trivial and  cannot have  
sections (i.e.   physical states) without vanishing points over
$\CM^\ast$ \cite{1}.
\qed
\enddemo

We have seen that physical states present two kind of nodes: those
located at reducible connections with
unstable bundles (Proposition 5.1) and the new ones required by 
Theorem 5.3 to appear at 
irreducible connections. The first type of nodes can be
thought as kinematical nodes, they appear for any physical state at
the same configurations because of its
geometrical origen based on the global properties of Gauss law.

However, the second type of nodes whose existence is inferred
from  theorem 5.3 are  genuine
non-abelian configurations and might depend on the  physical states
we consider. To find these irreducible nodal configurations a more
elaborate dynamical argument is required. In  general, these nodal
configurations will not be the same for all physical states. For
such a reason they can be considered as dynamical nodes.
 
It has been conjectured \cite{3}\ that the nodes of the vacuum
state $\Psi_0$, i.e. the eigenstate  of the hamiltonian (5.1) with
lowest eigenvalue, are located at connections which are associated to
unstable bundles. The
conjecture is based on the Ritz variational  argument which establishes
that the expectation value of the hamiltonian $\IH$ on
the  space of quantum
physical states must attaint its minimum value for the ground
state $\Psi_0$. It is trivial to see that the lowest eigenstates of
the  kinetic term of the hamiltonian (5.1) are the Chern-Simons
states discussed on the previous section \cite{2} and they vanish
for the strata associated unstable bundles. The contribution of the
potential term attaints its minimal value on these strata at the
reducible connections which satisfy the Yang-Mills equation
according to the Atiyah-Bott result \cite{8}. But the vacuum state
as any other physical state must vanish for such connections.
The vanishing of the vacuum wave functional $\Psi_0(A)$
along the whole strata seems,
thus, necessary to minimize the expectation value of the
hamiltonian $\IH$. This behavior is not only required for the
minimization of the kinetic term, but also for that of the
Yang-Mills potential term. Both terms, kinetic and potential, of the
hamiltonian conspire to force the vanishing of the vacuum 
on connections associated to unstable bundles
\cite{3}.

\remark{Remark\ {}\ }{Such a behavior is in contrast with Feynman's
claim on the absence  of nodes in the ground state of pure 
Yang-Mills theory in 2+1 dimensions \cite{12}. Another difference
between both theories is that in Yang-Mills theory fermions are
confined whereas in the topologically massive theory they are not.
This   relationship between the absence of confinement
and the existence of nodes in  the vacuum state, suggests that
those configurations where the vacuum vanishes might play a 
relevant role in the mechanism of  confinement.}
\endremark

 In compact lattice 
QED$_{2+1}$ it
has been  shown
that the logarithmic perturbative  Coulomb potential becomes linear
by means of Debye screening of electric charges in a monopole gas
\cite{27} in a similar manner as vortices drive the
Berezinskii-Kosterlitz-Thouless phase transition in the XY model
\cite{9}
 \cite{19}.
The above conjecture indicates that   connections
associated to unstable bundles would play a similar role in
the mechanism of quark confinement of the non-abelian theory.
This provides a geometric setting for the 't Hooft-Mandelstam
scenario \cite{18}\cite{22} in 2+1 dimensional gauge theories.

\remark{Remark\ {}\ }{ Although properly speaking,
there are not unstable bundles for infinite volumes ($\Sigma=\IR^2$), 
it is possible to distinguish  holomorphic bundles which can be extended
at infinity to stable bundles defined over the sphere from the ones
which do not admit such an extension. Actually, the bundles of the second
class  persist as nodes  of the quantum vacuum which suggests
that their contribution to the  confinement mechanism survives to
the infinite volume limit} \endremark

\head 6. Four-dimensional Gauge Theories.\endhead

The prominent relevance of unstable bundles in 2+1 dimensional
theories raises the question about their possible implications in 3+1
dimensional gauge theories.

Among the interesting classical configurations which have been
conjectured to have a role in non-perturbative effects in four
dimensions there are two leading configurations: instantons and
 sphalerons. Both solutions appear as classical solutions 
of Yang-Mills equations. 

{\it Instantons} are solutions of the
selfdual equations and therefore they are strict minima of the
Euclidean Yang-Mills functional in the corresponding
non-trivial bundle. Instantons with unit topological charge  
and structure group $SU(2)$ 
are given by
$$ A_\mu={2 \tau_{\mu \nu}
(x-x_0)^\nu\over (x-x_0)^2+\rho^2} 
\tag {6.1}$$ 
in stereographic coordinates of the four dimensional  sphere $S^4$ .
There are two collective coordinates which parametrize the moduli
space of $k=1$ $SU(2)$ instantons: the radius $\rho$ and its center
$x_0$. The ${\goth sl}(2,\IC)$  matrices $\tau_{\mu\nu}= i(
\tau_\nu^\dagger \tau_\mu - \delta_{\mu\nu})$, with 
$\tau=(-I,i\vec{\sigma})$ define a coupling between  internal and
external degrees of freedom.

Although the role of instantons 
as main responsible of the non-perturbative contributions associated
to tunnel effect between classical vacua seems to be very similar
to that of monopoles in compact QED in three-dimensional
space-times, their contribution to confinement does not seem to be
crucial. Indeed, a standard argument due to Witten shows that their
contribution is exponentially suppressed in the large N limit,
whereas quark confinement is strengthened in that limit 
\cite{30}. However, the instanton contribution is very relevant for 
the problem of chiral symmetry breaking in the presence of dynamical 
quarks \cite{10}.
 
{\it Sphalerons} are static solutions of Yang-Mills equations which can
only exist for finite space volumes. They are unstable
and become characterized by the existence of a finite number of
unstable decaying modes. The value of the Yang-Mill functional on
them marks the height of the potential barrier between classical
vacua and, therefore, is related to the transition temperature
necessary for the appearance of direct coalescence between those
vacua. In 2+1 dimensional gauge theories Atiyah and Bott showed that
in any strata of $\CA$ there exist sphalerons \cite{8}. They
correspond to the magnetic monopoles described in the previous
section, which where identified as kinematic nodes of all physical
states in topological massive gauge theory. In 3+1 dimensions their
existence was  predicted by  Manton \cite{23} by an argument based
on the  non-simply connected nature of the orbit space of 3
dimensional gauge  fields and a generalization of the
Ljusternik-Snierelman theory \cite{21}. Explicit expression for some
sphalerons on $S^3$ can be obtained from the observation that the
pullback of the instanton to a 3-dimensional sphere embedded on
$S^4$ with origin  on the center of the instanton and the same
radius that the instanton is an unstable critical point of the
3-dimensional Yang-Mills functional. For $SU(2)$  the sphaleron in
stereographic coordinates reads
$$ 
A_j={4\rho\over (x^2+4 \rho^2)^2}(4\rho \epsilon^a
_{jk} x^k - 2x^a x_j+ [x^2-4\rho^2]\delta^a_j)\sigma_a.
\tag{6.2}$$
 The unstable mode can be identified with the variation under
  scale transformations.
 
 The role of sphalerons in the quantum theory of gauge fields 
 comes from the following result.
\proclaim{Theorem 6.1}
 The vacuum state of the quantum 
 Yang-Mills theory  with a $\theta$--term ($\theta$--{vacuum})
has a node at any classical sphaleron 
 configuration for $\theta=\pi$.
\endproclaim
See \cite{4} for a proof. 

Since the theory is expected to deconfine
 for such a value of the theta parameter, the result suggest that those
 nodes might be responsible for the confining properties of the vacuum in
 absence of $\theta$ term where the vacuum has no classical nodal 
 configurations \cite{12}.

What is striking is that both types of 
connections, instantons and sphalerons, are related to unstable
bundles. If we consider the restriction of the sphaleron (6.2)
to a two dimensional sphere $S^2$ centered at (0,0,c)  and radius $a$
 we obtain in stereographic coordinates
the following two-dimensional connection
$$
A_{\bar z}= {\gamma\over
(1+|z|^2)^2}\pmatrix {- z} & -1\crcr z^2 & z\endpmatrix,
\tag{6.3}$$
 with 
$\gamma= 2a(a-ic)/(a^2+\rho^2 +c^2)$. It is easy to
show that such a connection corresponds to semi-stable bundle for
all values of $\gamma$, except for $\gamma=1$ where as shown Example
2.2 it corresponds to an unstable bundle. Such an
especial configuration emerges when the radius of the
embedded $S^2$ sphere
equals that of the  sphaleron $a=\rho$ and is centered at the origin 
$c=0$. This establishes 
a clear connection between sphalerons and unstable bundles.

In the same way it can be shown that instantons have a
correspondence with unstable bundles. If we considered the
embedding
of a 2-dimensional sphere $S^2$ of radius $a$ on $\IR^4$ 
with origin
at $x=(c,0,0,0)$. 
The pullback of the  instanton centered at $x=(0,0,0,0)$
to $S^2$ by such an
embedding is again given by (6.3) with 
$$\gamma={2a(a-ic)\over a^2+\rho^2+c^2 }. $$ 
Once more when and
only  $c=0$ and $a=\rho$ we get an unstable bundle, the same one
induced by the sphaleron.

This relation between instantons and unstable bundles can be 
generalized for instantons with higher topological charges. 
The  2-instanton configuration with two symmetric centers at
$x_+=(x_0,0,0,0)$ and $x_-=(-x_0,0,0,0)$ and one single scale $\rho$
reads,
$$ A_\mu=2 \overline{\tau}_{\mu \nu}
\partial^\nu \phi(x) $$
with
$$\phi(x)=\log\left(1 +{\rho^2\over (x-x_+)^2}+{\rho^2\over 
(x-x_-)^2}\right)
$$
and $\overline{\tau}_{\mu\nu}= i( \overline{\tau}_\nu^\dagger 
\overline{\tau}_\mu - \delta_{\mu\nu})$,
$\overline{\tau}=(-I,-i\vec{\sigma})$.
The corresponding pullback to a two
dimensional sphere centered at $x=(c,0,0,0)$ with radius $a$ is again
of the form (6.3) with    $$\gamma=-{a\over\rho^2} 
{2i\over 1+\alpha_+ +\alpha_-}\left[\alpha_+^2(c+x_0+ia) +
\alpha_-^2(c-x_0+ia)  \right] ,$$
with $\alpha_\pm=\rho^2/[x_0\pm c)^2+a^2]$. Therefore, in this case we
have two different spheres with unstable bundles. If the
distance between the two instanton centers is larger than the instanton 
size,
$\rho < 2 x_0$, the two spheres  have the same radius
$a\in(0,\sqrt{3}x_0)$  and are centered at two symmetric points 
$c_\pm=(\pm c(\rho,x_0),0,0,0)$ with $c(\rho,x_0)\in (0,x_0)$. When
the size of the instanton $\rho\downarrow 0$ the radius of the spheres
$a\downarrow 0$ and the centers go to $c_\pm\to x_\pm $. If the two
instantons are far apart $x_0\uparrow\infty$ the centers of the
spheres go to $x_\pm$ with radius $a=\rho$, as should correspond
to two isolated instantons. For instantons of larger radius $\rho> 2
x_0$ 
 the two spheres are centered on the origin
 with radii $$a_\pm=\sqrt{\rho^2-x_0^2\pm \rho\sqrt{\rho^2-4x_0^2}
}. $$ 
In the limit $x_0\downarrow 0$ when the distance between the 
instantons becomes negligible $a_+ \to \sqrt{2} \rho$ which agrees
with the fact that the two instantons merge into
a single instanton with radius $\sqrt{2} \rho$.

The same connection with unstable bundles appears for higher order 
multi-instantons. This is
not surprising because there are topological reasons which imply that
 generic multi-instanton connections must induce  unstable bundles
in some 2-dimensional spheres. 
 \proclaim{Proposition 6.2}
 A generic unitary connection $A$ on a principal bundle \hfill\break
$P(S^4,SU(N))$ 
 with non-trivial second Chern class
 $c_2(A)\neq 0$ induces  an unstable bundle over some 2-dimensional
 sphere $S^2$  embedded in $S^4$.
\endproclaim
\demo{Proof}
It is a straighforward generalization of the Asorey-Mitter result proved in
Ref. \cite{6} for connections over $S^2\times S^2$. 
 A generic unitary connection $A$ on a principal 
bundle $P(S^2\times S^2,SU(N))$ 
 with non-trivial second Chern class
 $c_2(A)=k$ induces by dimensional 
 reduction a non trivial 2-dimensional cycle $\sigma$ on 
 the orbit space of connections of the trivial bundle
 $S^2\times SU(N)$. In fact  $\sigma$ belongs to the $k$ 
 class  of second homotopy group of $\CM$, $\pi_2(\CM)=\IZ $
 \cite{6}. The orbits 
 of the connections
 induced by dimensional reduction are essentially defined by the 
 pullbacks of $A$ to a family of 2-dimensional spheres $S^2$ 
 embedded in $S^4$ (see Ref. \cite{6} for details). The 
generalization of the result for connections over the sphere 
$S^4$ is straightforward.

 Since the space of gauge orbits 
$\CM_{\hbox{\sevenrm ss}}$ 
induced from semi-stable bundles of
  $S^2\times SU(N)$ has a trivial cohomology group $\pi_2(
\CM_{\hbox{\sevenrm ss}})=0$
  any non-trivial cycle $\sigma$ on  $\pi_2(\CM^\ast)=\IZ $ must
cross some 
  of the
  orbits associated to unstable bundles.
  \qed
  \enddemo
  
However, the connection with unstable bundles is not exclusive of
4-dimensional gauge fields with non-vanishing $c_2$ Chern class. It
is trivial to find examples of connections belonging the classical
vacuum sector  with null Chern class, $c_2(A)=0$, which also induce
unstable bundles on some 2-dimensional spheres.

 \head 7. Conclusions\endhead
  
  We have analyzed the structure of  connections which appear as nodes
  of vacuum states in some quantum field theories. A common characteristic
  of all those connections is their relation with unstable bundles. Since
  in the corresponding quantum theories the quark interaction is shown to be 
  non-confining, it
  has been conjectured that there must exist a relation between those
  configurations and the mechanism of quark confinement in  gauge
  theories where such a nodes do not appear. On the other hand the fact that
  all those connections are in a way or another related with unstable 
  bundles suggests that it precisely this feature which is on the
  root of such a mechanism. To some extent  the
  association of connections with unstable bundles on  2-dimensional
  embedded surfaces can be considered as an intrinsic generalization 
  of the concept of magnetic monopoles
  for  non-abelian gauge theories. From this perspective the 
  above conjecture about the role of those configurations on the 
  confinement problem can be considered as a generalization of 't 
  Hooft-Mandelstam dual superconductor conjecture.


\refstyle{C}
 
\enddocument